# PythonPal: Enhancing Online Programming Education through Chatbot-Driven Personalized Feedback

Sirinda Palahan

*Abstract*— The rise of online programming education has necessitated more effective, personalized interactions, a gap that PythonPal aims to fill through its innovative learning system integrated with a chatbot. This research delves into PythonPal's potential to enhance the online learning experience, especially in contexts with high student-to-teacher ratios where there is a need for personalized feedback. PythonPal's design, featuring modules for conversation, tutorials, and exercises, was evaluated through student interactions and feedback. Key findings reveal PythonPal's proficiency in syntax error recognition and user query comprehension, with its intent classification model showing high accuracy. The system's performance in error feedback, though varied, demonstrates both strengths and areas for enhancement. Student feedback indicated satisfactory query understanding and feedback accuracy but also pointed out the need for faster responses and improved interaction quality. PythonPal's deployment promises to significantly enhance online programming education by providing immediate, personalized feedback and interactive learning experiences, fostering a deeper understanding of programming concepts among students. These benefits mark a step forward in addressing the challenges of distance learning, making programming education more accessible and effective.

## I. INTRODUCTION

THE challenges in online programming education, particularly for novice programmers, are significant and multifaceted. Students often struggle due to a lack of logical thinking processes, systematic problem-solving skills, and a deep understanding of algorithmic principles [1], [2]. Traditional teaching methods, which emphasize practical programming exercises, face hurdles as students grapple with syntactic errors and troubleshooting, a situation compounded by the high student-to-teacher ratios in many educational institutions, sometimes reaching 50:1 [3]. This scenario limits the effectiveness of personalized guidance and support crucial for learning programming.

The popularity of online programming courses has grown, offering wider accessibility but often missing the mark in providing personalized interaction and immediate feedback, both of which are essential in hands-on programming training. Innovations in online learning, such as peer review systems used by platforms like Coursera, do not always suffice for the specialized needs of novice programmers.

In the evolving landscape of online programming education, the integration of chatbots has revealed both promising potential and notable limitations. These AI-driven tools, emerging as early as the 1970s, are designed to simulate human interactions and have been incorporated into various educational settings, offering a semblance of personalized interaction akin to traditional classrooms [4]. Chatbots like "Coding Tutor" [5], "Sara, the Lecturer" [6], and "EduBot" [7], respond to programming queries and theoretical knowledge, yet they often lack in delivering hands-on coding practice and code-specific feedback, essential for practical application of theoretical concepts. Similarly, tools like "The Lecturer's Apprentice" [8] and the e-JAVA chatbot [9] offer programming exercises but do not evaluate student code, a key aspect in learning from mistakes.

Teaching programming poses unique challenges, requiring the effective conveyance of theoretical concepts and their application in coding exercises. Immediate, detailed feedback is critical in this context, as it bridges the gap between comprehension and application [10], [11]. Existing chatbots, while adept at addressing theoretical questions, frequently fall short in providing the level of feedback necessary for evaluating code and guiding practical exercises. This gap highlights the need for more specialized tools tailored to the unique challenges of programming education, capable of blending theoretical knowledge with practical coding exercises to offer a comprehensive learning experience. The rise of chatbots in education holds promise, but their effectiveness in domains like programming education necessitates enhancements to ensure they meet the specific needs of students grappling with new concepts and coding practices.

PythonPal is developed in response to the identified limitations within existing educational chatbots, aiming to significantly enhance online programming instruction. It merges theoretical lectures with practical exercises and incorporates interactive chatbot elements, thereby offering a balanced and

This paper was submitted on [date to be populated by IEEE]. This work was supported by University of the Thai Chamber of Commerce.

Sirinda Palahan is with the School of Science and Technology, University of the Thai Chamber of Commerce, Bangkok, 10400 Thailand (e-mail: sirinda_pal@utcc.ac.th).

immersive learning experience that bridges the gap between theory and practice. Unlike tools such as "Coding Tutor," which lacks comprehensive theoretical content, or "Sara, the Lecturer" and "EduBot," which focus more on programming queries than hands-on practice, PythonPal is designed to provide clear, natural language feedback on coding errors. This approach integrates advanced error message enhancement and interactive learning, ensuring that students not only engage in coding practice but also receive nuanced feedback essential for their development.

This research will evaluate PythonPal's effectiveness through two main research questions:

- **RQ1**: How accurate is PythonPal in identifying and providing feedback on common coding errors in programming exercises?
- **RQ2**: How do students perceive PythonPal's ability to comprehend queries, provide accurate feedback, and ensure overall satisfaction in the learning experience?

By addressing these research questions, this study aims to assess PythonPal's potential to revolutionize online programming education and foster the growth of well-equipped digital learners.

## II. LITERATURE REVIEW

In addressing the challenges of online programming education, this study primarily explores the technical capabilities of PythonPal, while recognizing the potential for future integration of social learning components such as student engagement, motivation, and collaboration. The literature review will examine both the technical capabilities of chatbots in programming education and the social dimensions in chatbot-mediated learning to provide a foundation for understanding PythonPal's current and potential impact on programming education.

### A. Technical Capabilities of Chatbots in Programming Education

Chatbots have emerged as powerful tools in the realm of programming education, offering a diverse range of functionalities. This literature review provides a comprehensive overview of their features and limitations in comparison to the PythonPal chatbot.

Hobert [5] introduced "Coding Tutor", a chatbot designed to assist students in learning Python programming. This chatbot primarily focuses on offering coding exercises and providing feedback based on the evaluation of student code. Hobert's chatbot guides students through Python programming tasks, emphasizing the syntax and semantics of commands. It boasts an architecture that includes adaptive learning paths, natural language processing for open-ended questions, and a source code analyzing component. However, while it effectively guides practical tasks, it lacks comprehensive theoretical content explanations. This limitation suggests that students might receive practical guidance but may miss out on deeper theoretical insights. Farah et al. [12] highlighted the challenges and opportunities of this approach. The chatbot in this study was designed to follow a rule-based script, offering explanations of Python programming best practices. However, its primary focus was on code style review rather than actual coding. The rule-based design, combined with declining student engagement over time, underscores the need for more intuitive chatbot tools that can maintain student interest.

On the other hand, some chatbots provide coding exercises but do not offer feedback on student code. For instance, "The Lecturer's Apprentice" developed by Ismail and Ade-Ibijola [8] serves as a virtual tutor, offering programming-based questions and solutions. Accessible via mobile phones, it aims to assist students, especially outside regular class hours. However, its lack of code evaluation means students do not receive feedback on their coding attempts. Similarly, the e-JAVA chatbot developed by Daud et al. [9] addresses basic Java learning content but doesn't evaluate or test student understanding. Such tools, while valuable, highlight the importance of feedback in programming education.

Transitioning from chatbots that provide coding exercises, there are those primarily designed to answer computer or programming-related questions without offering hands-on coding practice. Winkler et al. [6] introduced "Sara, the Lecturer," a chatbot that interacts with students during online instructional videos on basic programming. Sara's design is rooted in a scaffolding-based approach, tailoring guidance to a student's current understanding level. However, it primarily tests understanding of programming concepts and theories without offering practical coding exercises. Coronado et al. [2] proposed a modular cognitive agent tailored for e-Learning. This agent offers a conversational interface that serves as a quick reference guide. While it can elucidate programming concepts and even recommend associated topics, it doesn't provide practical coding exercises, potentially leading students to a superficial understanding of programming. Verleger and Pembridge [7] discuss the integration of "EduBot" an AI-driven chatbot, into an introductory programming course. EduBot was designed to offer real-time, instructor-moderated support. However, many students sought alternative resources when EduBot's responses were not immediate. EduBot's primary function is to answer questions and provide relevant information, but it doesn't offer direct coding instruction or exercises.

Some chatbots emphasize understanding coding concepts rather than hands-on coding. Benotti et al. [13] introduce the chatbot platform, which introduces high school students to Computer Science concepts through programming chatbots. Another tool, introduced in the paper by Fabic [14], is PyKinetic. It introduces a unique approach using Parson's problems paired with self-explanation prompts.

Another critical aspect of programming education is the feedback mechanism, especially when students encounter errors. Thiselton and Treude [15] introduce the PYCEE plugin, which augments Python compiler error messages by leveraging data from Stack Overflow. Another significant contribution in this domain is Paul Denny's study [16]. This research emphasizes the importance of crafting intuitive and inclusive error messages, especially for novice programmers. The insights from these tools can be invaluable for PythonPal. By integrating such advancements, PythonPal can offer more precise and understandable feedback messages, further enhancing the learning experience for students.

As the capabilities and limitations of current educational chatbots are explored, the influence of advanced AI technologies on their future development is recognized as increasingly critical, especially for personalized learning experiences and comprehensive feedback. Notably, the integration of AI-driven learning analytics stands out as an advancement with the potential to transform educational chatbots from generalized teaching assistants into personalized learning companions. AI-driven analytics could offer insights into student learning patterns, enabling chatbots to predict and address learning challenges before they become impediments [17], [18], [19]. These sophisticated analytics could empower chatbots like PythonPal to customize educational journeys for each student, tailoring content delivery, feedback, and challenge levels to individual learning styles and needs.

In conclusion, the literature underscores the innovations in educational chatbots and their unique contributions to programming education. Integrating AI-driven analytics emerges as a transformative step, making chatbots personalized learning aids that enhance education through tailored support. This shift towards customization highlights a broader trend in addressing educational needs, yet it also points to persisting challenges in programming instruction. A consistent observation is that while many tools excel in interactive teaching methods and feedback systems, there remains an evident need for comprehensive, context-specific feedback in coding education. This gap is especially crucial where understanding and correcting source code is concerned. PythonPal is designed to bridge this gap, drawing upon the insights from various studies focused on error message enhancement and interactive learning. By offering clear, natural language feedback on coding errors, PythonPal seeks to enhance students' error correction skills, thereby ensuring a more comprehensive and effective programming education experience.

*B. Social Dimensions in Chatbot-Mediated Learning*

While PythonPal focuses on technical precision in coding, it is also crucial to explore how chatbots, both in programming and non-programming contexts, integrate social learning components such as engagement, motivation, and collaboration. This section first reviews non-programming chatbots, many of which measure aspects of social dimensions like student engagement, motivation, and peer collaboration. These chatbots provide valuable insights into how social learning can be supported and evaluated in educational environments.

In text-based tutoring systems, including chatbots, feedback is crucial for improving engagement and motivation. For instance, example-tracing tutors using Cognitive Tutor Authoring Tools offer real-time, tailored feedback, highlighting the importance of timely, specific guidance in learning [20]. Studies show that automatic formative assessment increases engagement and task completion by providing immediate, corrective feedback. This fosters autonomy and supports long-term engagement [21]. Adaptive feedback systems support self-regulated learning by offering real-time guidance, helping students monitor and improve their progress, promoting resilience and success with complex tasks [22]. Tailored feedback also reduces cognitive load, making learning more efficient and motivating [23]. Overall, these findings stress the role of adaptive feedback and personalized content in creating effective learning experiences that benefit both technical mastery and student engagement.

Scaffolding within chatbot-student interactions also plays a key role in boosting motivation and confidence. Tailored scaffolding significantly enhances learning outcomes [24], guiding students step-by-step through complex tasks. These scaffolded conversations improve learning gains and task engagement [25], highlighting the value of chatbot-mediated interactions. Self-regulated learning is further enhanced by integrating metacognitive and motivational scaffolding through agents, with collaborative goal-setting leading to higher learning gains and better engagement [26].

Advancing beyond traditional one-on-one chatbots, multi-agent systems (MAS) offer enhanced educational experiences by facilitating both individual and collaborative learning. For example, self-adaptive MAS support real-time peer collaboration, fostering active learning environments [27]. Systems like I-MINDS and ClassroomWiki, which employ multiagent technology, further boost engagement and accountability in collaborative settings by organizing group learning, tracking contributions, and motivating students through feedback and peer comparison [28].

Social dimensions are measured through methods like interaction logs to track engagement, surveys to assess motivation, task completion rates [8], [20], [23] to evaluate persistence, and analysis of group contributions for collaboration [28]. These approaches provide insights into how student interactions, motivation, and persistence are enhanced by chatbot systems.

Having established the importance of social dimensions in non-programming chatbots, attention now shifts to programming chatbots, which were previously reviewed for their technical capabilities. These chatbots were primarily evaluated based on their effectiveness in enhancing the learning of programming concepts. While technical capabilities remain the central focus, it is also relevant to consider how the integration of social learning components contributes to the overall effectiveness of these chatbots.

In some papers, such as those discussing the e-JAVA Chatbot [6], PyKinetic [9], and Sara, the Lecturer [14], the focus is solely on the technical aspects of teaching programming, with no mention of social elements such as engagement or motivation.

Other papers, such as those on [2], [5], [7], [8], [12], and [20], acknowledge the importance of social dimensions like student interaction, motivation, and engagement. However, these aspects are often secondary to the main focus on technical integration. For instance, EduBot [7] and Code Style Bot [12] touch on student engagement but do not deeply explore these interactions. Lecturer's Apprentice [8] was designed to provide emotional support alongside academic assistance, yet its evaluation focused on general usefulness rather than measuring social impact. The chatbot in [2] incorporates social dialogue to enhance interaction, while [10] mentions motivation but without a detailed examination of its effects. Similarly, the Coding Tutor [5] acknowledges the potential of increasing student engagement through its interactive nature, but the emphasis remains on the technical aspects of learning to code.

Finally, some studies, such as those on a specific chatbot system [13], go further by measuring and evaluating the impact on student engagement, offering an understanding of how chatbots can enhance both technical learning and student interest.

The literature highlights the dual role of chatbots in educational settings: they not only enhance technical skills through real-time feedback and error correction but also support broader educational outcomes such as student engagement, motivation, and autonomy. PythonPal, with its focus on providing personalized and context-specific feedback, is well-positioned to bridge gaps in programming education by addressing both technical challenges and potentially enhancing student engagement through timely, actionable feedback. While PythonPal currently excels in enhancing technical skills, the integration of social dimensions—such as personalized scaffolding and adaptive feedback—presents a promising direction for future iterations. By combining technical precision with a deeper understanding of social learning components, PythonPal can evolve into a more comprehensive educational tool that supports both the technical and social aspects of learning.

## III. THE ARCHITECTURE OF PYTHONPAL

The system is bifurcated into two primary components: the chatbot interface and the PythonPal system, as depicted in Fig. 1. The chatbot interface facilitates user interaction, utilizing the LINE messenger app, Thailand's predominant messaging platform [29], ensuring wide-spread accessibility for students.

PythonPal is architecturally constructed with a focus on modularity, enabling seamless adaptability to a variety of tutorial sources, evaluation methods, and models. It encompasses three main modules: Conversation, Tutorial, and Exercise Modules. The Conversation module orchestrates student interactions in English, guiding them to either tutorial videos or exercise materials based on their inquiries. The Tutorial module serves as a reservoir for instructional videos. In contrast, the Exercise module supervises the submission, assessment, and feedback of student exercise solutions, accommodating a range of evaluation methodologies.

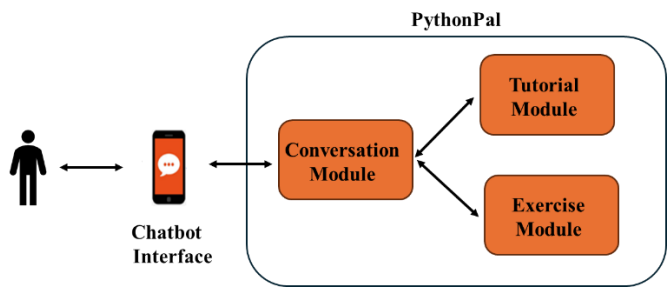

Fig. 1. The architecture of PythonPal.

### A. The Conversation Module

The Conversation module serves as an integral component of the system, designed to assist students in English. Beyond disseminating personalized lecture videos or exercise materials, it incorporates a chatbot. This chatbot employs an intent classifier to discern user intentions, be it seeking a video, an exercise, submitting an exercise, or merely engaging in a chat. When users opt for a conversation, a deep learning model facilitates the dialogue. Subsequent sections will explore the chatbot's conversational design, elucidate the intent classification process, and discuss the deep learning model underpinning these interactions.

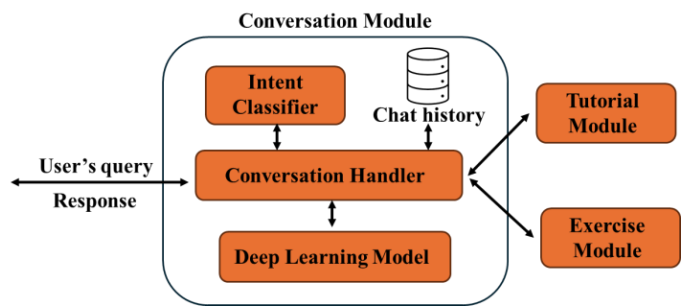

Fig. 2. Components in the Conversation Module.

*Conversation Design*

As depicted in Fig. 2, the Conversation module comprises several components. Central to this is the Conversation Handler, crafted to enhance the user experience by adeptly addressing their needs.

The process works like this:

- Input Acquisition: Users begin by entering their query.
- Intent Detection: Assisted by the intent classifier, the Conversation Handler endeavors to ascertain the user's intent.
- Request Routing & Handling: Based on the identified intent, the system offers an appropriate response, which could involve directing users to a tutorial, presenting an exercise, or initiating a chat.
- State Management: The chatbot maintains a record of recent user activities, especially about the exercise and tutorial modules they've accessed. For instance, if a user ex-presses a desire to submit an exercise, the chatbot inquires if it pertains to the most recent exercise they engaged with.
- State Update: While the chatbot doesn't retain entire conversation histories, it is designed to remember ongoing or pending tasks. For example, if a user was in the process of submitting an exercise, the chatbot would recall that context and guide the user accordingly.

Guided by the Conversation Handler, the chatbot is designed to offer a transparent and straightforward user experience. The aspiration is for users to find this interface both intuitive and beneficial. A cornerstone of this seamless experience is the chatbot's ability to accurately discern user intent during each interaction.

*Intent classification*

Understanding user intent is crucial. In the design of the chatbot, four main categories of student interactions were identified: 1) General conversation, 2) Requesting a video tutorial, 3) Requesting an exercise, and 4) Submitting an exercise. To cater to the specific learning needs of the students, six key topics within the realms of video tutorials and exercises were identified. These topics are: 1) Creating and Using Variables, 2) Data Types, 3) Basic Arithmetic Operations, 4) Comparison Operators, 5) Defining and Calling Functions, and 6) Function Return Values.

Based on these categories and topics, sixteen intents for the chatbot were defined. These include 'General conversation', 'Submitting an exercise', six distinct intents for 'Requesting a video tutorial' (one for each topic), and six distinct intents for 'Requesting an exercise' (one for each topic), one intent for a non-existent topic under 'Requesting a video tutorial', and one intent for a non-existent topic under 'Requesting an exercise'. In the hierarchical classifier, the first level of classification categorizes a student's query into one of the four main intents mentioned above. If the intent is either 'Requesting a video tutorial' or 'Requesting an exercise', the classifier then further categorizes the query into one of the six topics or an additional 'non-existent topic' category, resulting in a total of seven sub-intents. In contrast, the flat classifier directly categorizes a student's query into one of sixteen possible intents. This design allows for a comparison of the effectiveness of flat and hierarchical classification strategies in handling student queries in the chatbot system.

In the intent classification process, both flat and hierarchical classification strategies were examined. The comparison between these two strategies is essential due to their distinct approaches to classifying intents. Flat classification considers all classes equally, without any hierarchical structure, which can be efficient when dealing with a smaller number of classes. On the other hand, hierarchical classification leverages the inherent structure of classes, which can be particularly beneficial when dealing with a larger number of classes. By recognizing and utilizing the relationships between classes, hierarchical classification has the potential to enhance the accuracy of intent classification.

Several classification models known for their computational efficiency and suitability for text data were selected:

1) Support Vector Machines (SVM) [30]: SVMs are effective in high-dimensional spaces, such as text data, and versatile due to the different Kernel functions that can be specified for the decision function. Both a flat SVM model and a hierarchical SVM model were tested.
2) XGBoost [31] is an optimized distributed gradient boosting library. Both a flat and a hierarchical XGBoost model were tested.
3) DistilBERT [32] is a transformer model that maintains a high level of accuracy. It was used in a flat classification model.
4) FastText [33] is a library for efficient learning of word representations and sentence classification. It was also used in a flat classification model.

For both SVM and XGBoost models, three feature extraction methods were explored: tf-idf, word2vec, and GloVe:

1) TF-IDF (Term Frequency-Inverse Document Frequency) [34]: This method reflects how important a word is to a document in a collection or corpus.
2) Word2Vec [35]: This is a group of related models used to pro-duce word embeddings.
3) GloVe (Global Vectors for Word Representation) [36] This is an unsupervised learning algorithm for obtaining vector representations for words.

*Deep Learning Model for Conversation*

The Conversation module leverages deep learning models to enhance user experience, especially when managing general chat-like interactions. When users express the intent for a general conversation, the system resorts to a specialized deep-learning conversation model. Several models known for their proficiency in processing and generating human-like textual responses were explored. These models include:

1) GODEL [37]: Microsoft Research's Goal-Oriented Dialogue Learning model is adept at generating responses rooted in external texts. With its pre-training on a vast dataset, GODEL can be quickly adapted to specific dialogue tasks using limited task-specific data.
2) GPT-2 [38]: Developed by OpenAI, GPT-2 is notable for generating natural text, enhancing the conversational quality of the chatbot.
3) T5 [39]: The Text-to-Text Transfer Transformer model approaches every NLP challenge as a text-to-text problem, making it versatile for varied tasks.
4) BlenderBot [40]: Created by Meta, this model is tailored for open-domain interactions, emphasizing engaging and meaningful exchanges.

Models such as GODEL, GPT-2, T5, and BlenderBot have the advantage of being pre-trained on extensive text datasets. They can then be fine-tuned to cater to specific needs, such as the casual interactions in the PythonPal system. The eventual selection from these models will be based on their experimental performance. Factors under consideration will include the quality of responses (evaluated using METEOR and BLEU scores) and the efficiency of text data processing. The aim is to select a model that effectively combines high response quality with reasonable processing speed for the PythonPal chatbot.

### B. The Tutorial Module

The Tutorial module is structured around two main components: the Tutorial Handler and the Tutorial Repository as shown in Fig. 3

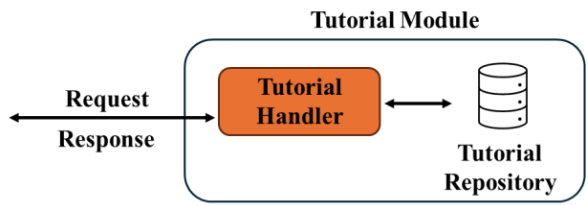

**Fig. 3.** Components in the Tutorial Module.

Upon receiving a tutorial request from the Conversation module, the Tutorial Handler takes charge. It's tasked with fetching the appropriate tutorial link and dispatching it back to the user. While the current design has the intent classifier within the Conversation module determining the tutorial topic a user is interested in, the future vision entails the Tutorial Handler assuming a more proactive role. The aim is for it to not only handle retrieval but also search and match the most fitting tutorial in response to a user's specific query.

Meanwhile, the Tutorial Repository is where the vast array of educational resources is housed. Predominantly, it stores tutorial videos touching on foundational Python programming topics like variables, basic arithmetic operations, and functions. These topics were chosen due to their inherent importance in understanding the core of Python. Notably, all these videos were produced by the team, ensuring a consistent level of quality and pedagogy. To maintain the exclusivity of the content, the videos are hosted on YouTube as unlisted, making them accessible only through direct links. The adopted repository method ensures systematic organization and efficient retrieval of these resources. As progress is made, the vision is to enrich the repository with a broader range of tutorial topics and continually refine the storage and retrieval mechanisms for greater efficiency.

### C. The Exercise Module

The Exercise module comprises two primary components: the Exercise Handler and the Exercise Grader, as illustrated in Fig. 4. When a request is received from the Conversation module, the Exercise Handler acts as the main routing mechanism. It interprets the intent conveyed from the Conversation module and initiates the corresponding action. If the intent is to request an exercise, the handler retrieves the link to the desired exercise from the Exercise Repository and returns it to the user. Conversely, if the intent pertains to the submission of an exercise solution, the handler directs the solution to the Exercise Grader for evaluation.

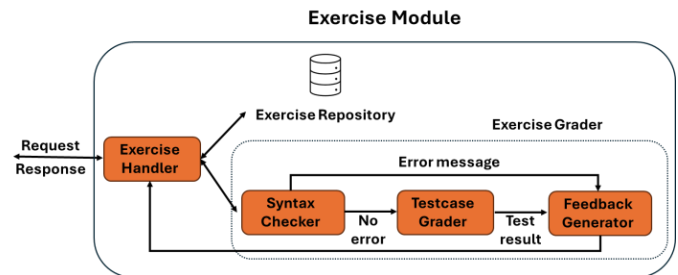

**Fig. 4.** Components in the Exercise Module.

Central to the evaluation mechanism, the Exercise Grader integrates three sub-modules: the Syntax Checker, the Testcase Grader, and the Feedback Generator. The Syntax Checker scrutinizes the student's code for syntactic accuracy. If the code clears this phase, the Testcase Grader examines the code's functionality using predefined test cases. Subsequently, the Feedback Generator formulates feedback messages based on the outcomes from the Syntax Checker and the Testcase Grader. These messages are relayed to the Conversation module, which then forwards them to the students, offering an assessment of their code.

Subsequent subsections delve into the detailed functionalities of each sub-module within the Exercise Grader, elucidating their roles and significance within the overarching PythonPal architecture.

*The Syntax Checker*

In devising an automated mechanism to pinpoint errors in students' code, principles from diverse fields, including computer science, software engineering, and programming education, were integrated. The following are pivotal theories and concepts incorporated:

1) Syntax Analysis: Syntax encompasses the rules defining the structure and composition of code in a specific programming language. A significant portion of errors students commit while coding pertains to syntactic mistakes, such as omitted parentheses or incorrect indentation. The Syntax Checker is tailored to detect and highlight these errors, aiding students in rectifying their code and grasping the significance of accurate syntax in programming.

2) Common Error Patterns: Historically, educators and scholars have discerned recurring error patterns among coding students [41]. These can encompass misconceptions of particular language features, confusion between analogous concepts, and logical errors. These patterns have been embedded into the Syntax Checker to anticipate errors based on context, amplifying its capability to identify and highlight potential issues in students' code.

3) Error Classification in Computer Programming: Established frameworks for categorizing programming errors were also consulted. A seminal taxonomy is attributed to Soloway and Ehrlich [42], which delineates error categories such as "plan composition errors," "plan instantiation errors," and "plan integration errors." This framework has been pivotal in discerning the types of errors to detect in student code.

An analysis of six core Python programming exercises revealed several common error patterns, which led to the development of a code snippet for error detection and clarification. The errors identified include:

1) Undefined Variable or Name Errors: These include missing string quotes, incorrect assignment operators, typographical errors in variable names, use of undefined variables, incorrect variable names, case sensitivity issues, and sequencing problems.
2) Typographical Errors: Common issues are end-of-line (EOL) errors in string literals, fundamental syntax mistakes, misuse of string format functions, and typographical errors in names.
3) Used Before Assignment Errors: Errors such as using variables before they are assigned, printing unassigned variables, and undefined variables in function calls.
4) Data Type or Value Errors: These encompass inappropriate type assignments, division by zero, incorrect data types in arithmetic operations, and unquoted strings.
5) Function-related Errors: Issues include premature function calls, incorrect call syntax, misunderstandings in function definition or invocation, return statement problems, and incorrect argument sequencing.
6) Indentation Errors: Problems arising from incorrect indentation, which is crucial in Python.
7) Logical or Semantic Errors: These include hardcoding answers, incorrect operator uses, hardcoded values within functions, incorrect operation sequencing, and operator misuse.

A pivotal aspect of error detection lies in the ability to present students with informative and constructive error messages. If an error does not align with a predefined pattern, the system defaults to displaying Python's original error message, which includes the error type and the line number where the error was detected. This approach offers students guidance on potential problem areas within their code. Furthermore, recognizing error patterns is an ongoing, evolving endeavor. As the system evaluates more exercises, novel error patterns may surface. Similarly, introducing new exercises to the curriculum might introduce unique errors not currently accounted for. Regularly updating the error detection module to accommodate these emerging patterns ensures the system's continued efficacy in both error identification and constructive feedback provision.

*The Test-case Grader*

Once the Syntax Checker has confirmed that a student's code is devoid of syntax errors, the Testcase Grader undertakes the subsequent evaluation phase. Each exercise is associated with a singular, general test case that the grader uses to assess the accuracy of the student's solution. This test case is not intricate in the context of software testing with multiple edge cases but is a direct test consistent with the introductory nature of the course.

The choice to integrate only one test case per exercise is influenced by the foundational nature of the content. Given the beginner-oriented design of this course, introducing learners to multiple test scenarios might be overwhelming and counterintuitive. Instead, the emphasis is on understanding basic concepts rather than managing every potential input variation. Moreover, to promote self-testing and comprehension, both the test case and its anticipated output are shared with the students. This transparency enables them to autonomously verify their code against the expected result before submitting.

*The Feedback Generator*

The Feedback Generator serves a dual purpose in the system. If syntax errors are identified in the student's code, it offers a comprehensive error message detailing the error type, its position in the code, and potential origins. This feedback aims to be lucid and instructive, guiding students in understanding the issue and its potential resolution. Conversely, if the student's code clears the syn-tax check, the Feedback Generator then comments on the Testcase Grader's results, indicating whether the code has met or failed the test case criteria, offering crucial insights into the code's functionality and accuracy.

Fig. 5 shows an example code that contains an error. It is used to illustrate and compare the feedback provided by PythonPal with the original Trace message. In Fig. 6, a comparison is made between the original feedback message (shown on the left) and the feedback message generated by PythonPal (shown on the right). The figure highlights the differences and improvements introduced by PythonPal's feedback system.

```
def average(a, b):
    s = (a+b)/2
    return s
average(2,6)
print (s)
```

**Fig 5.** An example code with an error.

| Traceback (most recent call last): | Error: Undefined variable 's' |
|---|---|
| File "C:\Users\t2.py", line 6, in <module> | Location: line 6 |
| print (s) | The error can be from: |
| NameError: name 's' is not defined | 1. You could use a variable name that's not been defined. (You may defined it after you use it) |
| | 2. There might be a typo in the variable name. |
| | 3. Check when you defined the variable that it is in the format: variable_name = value |

**Fig.6.** Comparison of Feedback Messages: Original (left) vs. PythonPal (right) for the code in Fig. 5.

*D. The Chatbot Interface*

In this subsection, we delve into the user interface aspect of PythonPal, which plays a crucial role in facilitating user interaction with the system's sophisticated backend. The interface is designed to be intuitive and user-friendly, allowing students to navigate through various functionalities seamlessly.

The chatbot interface of PythonPal handles a range of student interactions, from simple queries about programming concepts to requests for specific exercises or tutorials. Fig. 7 serves as a visual guide. It demonstrates the chatbot's versatility in handling different types of student interactions. For instance, Fig. 7a captures a general conversation between a student and the chatbot. Fig. 7b and 7c illustrate the chatbot's responses to requests for a video tutorial and an exercise on basic calculation and calling functions, respectively. Finally, Fig. 7d presents the feedback given on a student's submitted work.

## IV. THE METHODOLOGY OF THE EXPERIMENTAL DESIGN

The goal of this experimental design is to evaluate the effectiveness of PythonPal's three core modules - the Conversation, Tutorial, and Exercise modules - in terms of functionality and user satisfaction. To achieve this, four experiments are designed, each targeting specific aspects of PythonPal:

1) Intent Recognition in Conversation Module: This experiment evaluates the Conversation module's ability to classify and interpret user queries accurately. A variety of user queries are used to assess the module's effectiveness in recognizing intents. The performance is measured using the F1-score, chosen for its ability to balance precision and recall, thereby providing a more holistic view of the module's accuracy. Additionally, inference time is measured, offering insights into both the accuracy and efficiency of the module.

2) Conversational Coherence in Conversation Module: This experiment concentrates on enhancing and evaluating the Conversation module's ability to maintain context and coherence in dialogues. Deep learning models are applied and fine-tuned for this purpose. Their effectiveness is tested in simulated dialogues, and the conversation quality is assessed using METEOR and BLEU scores, alongside inference times. METEOR and BLEU scores are used as they are established metrics in natural language processing, ideal for gauging the linguistic accuracy and fluency of the chatbot's responses.

3) Exercise Module Accuracy in Error Feedback: This section addresses Research Question 1 (RQ1) - "How accurate is PythonPal in identifying and providing feedback on common coding errors in programming exercises?" It focuses on evaluating the Exercise module's ability to detect coding errors and provide accurate feedback. The experiment involves using programming exercises embedded with common errors, and the feedback provided by the module is assessed for accuracy using an accuracy rate metric.

4) User Satisfaction Evaluation: Aligned with Research Question 2 (RQ2) - "How do students perceive PythonPal's ability to comprehend queries, provide accurate feedback, and ensure overall satisfaction in the learning experience?" - this experiment assesses user satisfaction across all modules. The methodology includes user experience testing with surveys and feedback collection, utilizing Likert scale ratings and qualitative analysis for a thorough evaluation of user satisfaction.

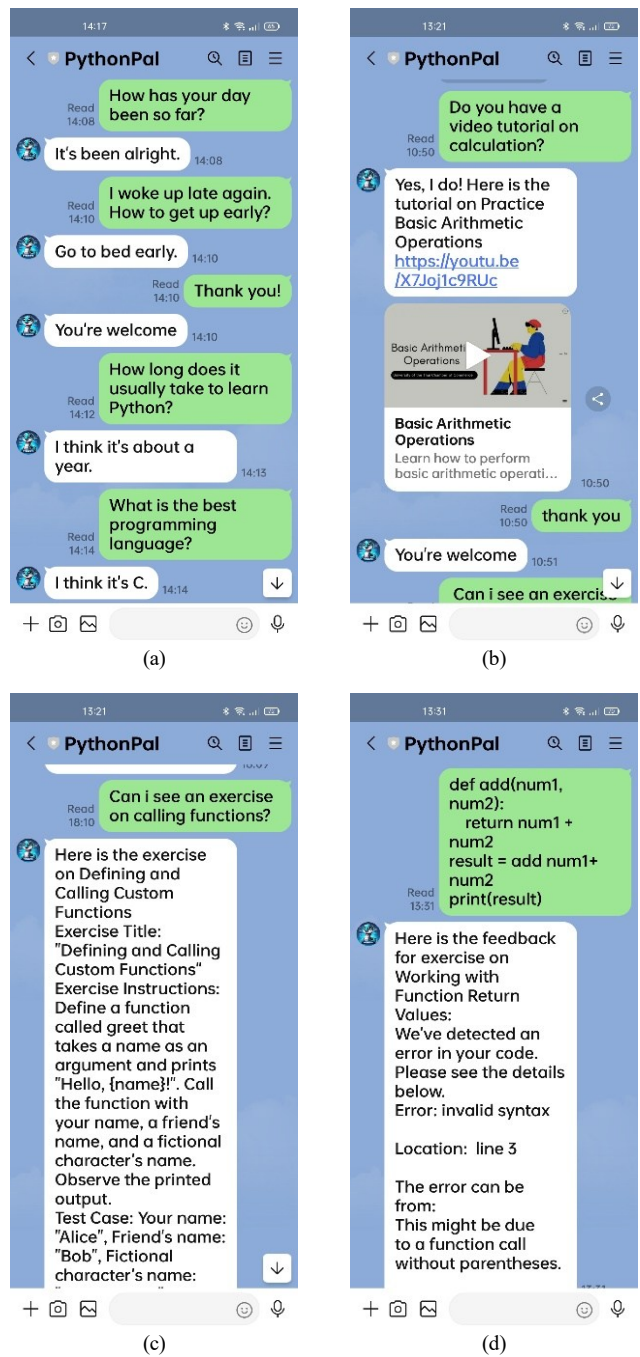

**Fig. 7.** Chatbot responses to a student's query: a) A general conversation b) Request a video c) Request an exercise and d) Request feedback.

## V. EXPERIMENTAL RESULTS AND DISCUSSIONS

The experimental segment of this research aims to offer an exhaustive assessment of PythonPal. This assessment is organized around four pivotal elements, each playing a significant role in determining the overall efficiency and user-friendliness of the chatbot. Subsequent sections will explore these essential facets in detail.

### A. Intent Recognition in Conversation Module

The capacity for precise user intent classification is foundational to the efficacy of any chatbot. This segment explores the processes that empower PythonPal to adeptly discern user inquiries.

*The Dataset*

The dataset was curated to reflect the diverse range of queries the system is likely to encounter. It encompasses 800 inquiries, uniformly spread across different intents, with each intent having 50 queries. Such equitable distribution guarantees that the classifiers are educated on a sample mirroring the varied inquiries PythonPal might face. The dataset was created by a team of research assistants, all of whom are undergraduate students with backgrounds similar to the survey's participant pool. This selection was intentional, as it aligns the queries' style and complexity with the expected user base of PythonPal. The research assistants were instructed to formulate queries that they or their peers might typically pose while learning Python, thereby capturing a realistic and relevant range of student inquiries.

To ensure the dataset's accuracy and relevance, several key measures were implemented. The intent of each query was verified by the researcher, ensuring accurate alignment with the expected use cases of PythonPal and maintaining the dataset's integrity. Consistency checks were regularly conducted across the dataset, where queries among different intents were cross-verified. This process ensured a uniform standard of complexity and relevance, mitigating any biases or inconsistencies. For an in-depth examination of the query types and their associated intents, see Table I.

*The evaluation results*

Table II showcases the experimental outcomes for diverse models, features, and evaluation metrics. Each model's performance is gauged using the F1-score and inference duration per query (in seconds). Within the hierarchical models, the SVM equipped with GloVe features registered the peak F1-score at 0.8555, signifying proficient intent classification. However, this model exhibited a protracted training duration in contrast to the SVM with TF-IDF features. The XGBoost model, when paired with TF-IDF features, also demonstrated commendable performance, registering an F1-score of 0.8256, and boasted the briefest inference time among its hierarchical counterparts. Regarding the flat models, the SVMs with TF-IDF, Word2Vec, and GloVe features all recorded comparable F1-scores around 0.895, with the TF-IDF feature marginally surpassing its peers. The XGBoost model, when combined with TF-IDF features, also showcased commendable performance, achieving an F1-score of 0.8894. It's noteworthy that flat models generally exhibited swifter inference times relative to hierarchical models.

TABLE I
EXAMPLES OF QUERIES AND THEIR CORRESPONDING INTENTS

| Query | Intent |
|---|---|
| How can I stay motivated while learning to code? | general_conversation |
| Can you show me a video on Python variables? | video_variable |
| Can you show me a video on understanding data types in Python? | video_data_types |
| Can I watch a video on using Python for arithmetic calculations? | video_arith_operation |
| I need a video on defining and calling functions in Python. | video_functions |
| Can you find a video on how to work with return values in Python functions? | video_return_values |
| Please provide a beginner's exercise on Python data types. | exercise_data_types |
| I'd like to practice basic arithmetic in Python. | exercise_arith_operation |
| I need a practice exercise on defining and calling functions in Python. | exercise_function |
| Is there a video explaining the difference between global and local variables in Python? | video_none |
| I'd like an exercise about calculating the power of a number in Python. | exercise_none |
| I've finished the exercise and I'm ready to submit. | ask_submission |

In the realm of transfer learning models, DistilBERT outshined its counterparts, registering the pinnacle F1-score of 0.9352, underscoring its excellence in intent classification. Nonetheless, it exhibited an extended training and inference duration relative to FastText, which, while achieving a commendable F1-score of 0.8315, boasted the swiftest inference time across all models.

Of all models evaluated, DistilBERT stood out as the premier performer, achieving the zenith F1-score of 0.9352, which underscores its superior intent classification capabilities. Even though its training and inference durations were on the lengthier side, DistilBERT's precision resonates with the primary objective of delivering accurate replies. The significance of prompt responses is also recognized. Consequently, forthcoming endeavors will center on enhancing DistilBERT's efficiency without undermining its precision, potentially employing strategies such as model distillation or pruning.

TABLE II
EXPERIMENTAL RESULTS OF INTENT CLASSIFICATION MODELS PERFORMANCE

| Model Type | Model | Feature | F1-score | Inference time(s) |
|---|---|---|---|---|
| Hierarchical models | SVM | TF-IDF | 0.8226 | 0.000374 |
| | | Word2Vec | 0.8088 | 0.000177 |
| | | glove | 0.8555 | 0.000145 |
| | XGBoost | TF-IDF | 0.8256 | 0.003103 |
| | | w2v | 0.8024 | 0.002860 |
| | | glove | 0.7447 | 0.002773 |
| Flat models | SVM | TF-IDF | 0.8982 | 0.000063 |
| | | w2v | 0.8950 | 0.000108 |
| | | glove | 0.8956 | 0.000063 |
| | XGBoost | TF-IDF | 0.8894 | 0.000040 |
| | | w2v | 0.7999 | 0.000027 |
| | | glove | 0.7601 | 0.000026 |
| Transfer learning | DistrilBERT | | 0.9352 | 0.001402 |
| | FastText | | 0.8315 | 0.000016 |

*B. Conversational Coherence Enhancement in the Conversation Module*

Building upon the foundational understanding of user intent classification, this section emphasizes the deep learning models that underpin PythonPal's conversational capabilities. The fine-tuning process is detailed, utilizing the DailyDialog dataset to ensure the chatbot's interactions remain contextually appropriate and natural.

*The Dataset*

The DailyDialog dataset [43] was chosen for this endeavor, a comprehensive collection of 13,118 multi-turn dialogues in English covering a range of everyday topics. These dialogues not only mirror typical conversational patterns but are also categorized based on specific themes and settings. To resonate with the educational context of PythonPal, dialogues from the 'School life' category were selected, encompassing 487 dialogues. This selection aimed to prepare the chatbot for conversations it might frequently encounter within an academic environment.

An example of a dialogue from this topic is as follows:

```
"What do you think of the test ? __eou__ Rather difficult . How about you ? __eou__ It wasn't too bad . Were you nervous ? __eou__ Rather . I'm afraid I didn't do very well . __eou__ Oh , I wouldn't worry about it . __eou__ That's because you did well . __eou__ No , I wouldn't worry about you . __eou__ Why ? __eou__ It seems to me that you always do better than you say . __eou__
```

Here, `__eou__` signifies a change in speaker turn. To prepare the data for model training, pairs of turns for each dialogue were created. For instance, ('What do you think of the test ?', and 'Rather difficult.') would be one such pair. In total, 4,084 pairs were generated. These pairs were then divided into training, validation, and test datasets using a 60/20/20 split. This allocation ensures that the models receive training on a diverse set of dialogues and are evaluated on data they have not previously encountered, offering a reliable assessment of their capabilities.

*The Evaluation Results*

Four pre-trained models were fine-tuned: GODEL, GPT-2, T5, and BlenderBot. These models, accessible on the Hugging Face platform, have undergone pre-training on extensive text corpora, making them apt for the task of generating conversational responses. The fine-tuning was executed on Google Collab, leveraging the capabilities of a V100 GPU. Each model underwent training on the 'School life' dialogue dataset for three epochs. During this phase, the training time for each model was observed and documented, shedding light on their computational efficiency. After training, the models were evaluated based on their ability to predict a response given a source input. This involved measuring the inference time, which denotes the duration the model takes to produce a response.

To assess the quality of the generated responses, two evaluation metrics were computed: METEOR and BLEU scores on the test dataset. These metrics offer a quantitative evaluation of the models' proficiency in producing responses that align with the target responses in the dataset.

TABLE III
EXPERIMENTAL RESULTS OF DEEP LEARNING MODELS PERFORMANCE

| Model | METEOR | BLEU | Inference time (s) |
|---|---|---|---|
| GODEL | 0.1056 | 0.0226 | 0.2333 |
| GPT-2 | 0.1136 | 0.0014 | 0.0276 |
| T5 | 0.1035 | 0.0185 | 0.2076 |
| BlenderBot | 0.1504 | 0.0019 | 0.0978 |

Table III offers a detailed comparison of performance and efficiency metrics for the four models assessed in the experiments: GODEL, GPT-2, T5, and BlenderBot. Each model's evaluation is based on the METEOR score, BLEU score, and inference time per response. Although BLEU scores are also presented, they mainly validate the experimental setup rather than act as a primary selection criterion. The BLEU scores from these experiments align with those presented in the study by Li et al. [43] reinforcing the credibility of the experimental design.

Examining the METEOR scores, which reflect the model's response quality in terms of relevance, BlenderBot is identified as the superior model with a score of 0.1504, markedly outperforming the other models. GPT-2 is next with a METEOR score of 0.1136, while GODEL and T5 present slightly lower scores of 0.1056 and 0.1035, respectively. Regarding efficiency, GPT-2 excels with the fastest inference time at 0.0276 seconds, making it highly suitable for real-time interactions in PythonPal. BlenderBot, with a respectable inference time of 0.0978 seconds, also emerges as a viable option, offering a balance between speed and performance. While T5 is a bit slower at 0.2076 seconds, both GPT-2 and BlenderBot stand out as strong candidates for PythonPal, particularly for scenarios demanding quick feedback.

After considering the METEOR scores and inference times of GODEL, GPT-2, T5, and BlenderBot, BlenderBot is recognized as the optimal model due to its outstanding METEOR score of 0.1504. While its inference time per response is not the quickest, it remains within a tolerable range and has the potential for enhancement with increased computational resources. As a result, BlenderBot offers an impressive combination of top-tier dialog generation and acceptable efficiency, positioning it as the preferred choice for the system.

*C. Exercise Module Accuracy in Error Feedback*

This section evaluates PythonPal's Exercise Module, addressing Research Question 1 (RQ1): How accurate is PythonPal in identifying and providing feedback on common coding errors in programming exercises? The focus is on assessing PythonPal's ability to detect and correct coding errors, using a specially created dataset that mimics real student challenges in programming.

*The Dataset*

To establish an unseen dataset for the evaluation of the chatbot system's accuracy, expertise was sought from three lecturers affiliated with the Department of Information and Communication Technology and the Department of Computer Science. Each lecturer was tasked with crafting two error-prone solutions for each of the six exercises. This method resulted in 6 solutions per exercise, amassing a total of 36 unique solutions. Leveraging their expertise, the lecturers were guided to incorporate common errors that novices frequently make during their programming learning journey. However, they were not given a specific list of detectable errors by the system. This strategy was employed to assess the system's proficiency in identifying and providing feedback on prevalent coding errors. It is also pertinent to mention that these lecturers had no prior involvement in the chatbot's creation and were unfamiliar with the specifics of its error detection capabilities, ensuring an unbiased approach to the solutions they crafted. Importantly, these lecturers did not participate in evaluating PythonPal itself; their contribution was solely to create a dataset for testing PythonPal's error detection and feedback mechanisms. Their work aimed to enhance PythonPal's precision in identifying errors, indirectly suggesting that such improvements could boost user satisfaction by providing accurate feedback, essential for effective learning and understanding of programming concepts.

*The Evaluation Results*

The evaluation aimed to gauge the system's performance, with a primary focus on accuracy, an essential metric for interactive learning tools. Within this framework, accuracy denotes the proportion of solutions where the system accurately identified an error and offered relevant feedback. This metric evaluates the system's proficiency in detecting coding mistakes in students' submissions and providing precise feedback on those errors. The assessment employed a dataset of unfamiliar, error-prone solutions developed by educators, as outlined in the Dataset section. The evaluation results are presented in Table IV, offering a detailed view of the system's performance in identifying and addressing coding errors within these solutions.

TABLE IV
ACCURACY EVALUATION ON UNSEEN SOLUTIONS

| Exercise Title | Accuracy |
| --- | --- |
| Variable Creation and Assignment | 100% (6/6) |
| Exploring Python Data Types | 83.33% (5/6) |
| Practice Basic Arithmetic Operations | 100% (6/6) |
| Comparing Values with Comparison Operators | 66.67% (4/6) |
| Defining and Calling Custom Functions | 83.33% (5/6) |
| Working with Function Return Values | 83.33% (5/6) |

To gain a deeper understanding of potential areas of improvement for PythonPal, the solutions that the system either failed to recognize or provided inaccurate feedback were analyzed. This analysis is presented in Table V. In this table:
- The "Incorrect Feedback" column represents the number of solutions where the chatbot detected an error but provided an incorrect explanation.
- The "Undetected Errors" column represents the number of solutions where the chatbot failed to detect or recognize an error in the student's code, resulting in incorrect feedback.

TABLE V
ERROR DETECTION AND IDENTIFICATION RESULT

| Exercise Title | Incorrect Feedback | Undetected Errors |
| --- | --- | --- |
| Variable Creation and Assignment | 0 | 0 |
| Exploring Python Data Types | 0 | 1 |
| Practice Basic Arithmetic Operations | 0 | 0 |
| Comparing Values with Comparison Operators | 0 | 2 |
| Defining and Calling Custom Functions | 0 | 1 |
| Working with Function Return Values | 1 | 1 |

From Table V, it becomes apparent that the chatbot performs well with exercises such as "Variable Creation and Assignment", where no errors went undetected or explained incorrectly. However, there is one solution that the chatbot provided incorrect feedback. In the exercise "Working with Function Return Values", the chatbot correctly pinpointed the error and its location at:

```
def add(num1, num2)   # Missing colon at the end
    return num1 + num2
```

The actual cause of the error was the missing colon at the end. However, the system mistakenly explained it as potentially being due to a function call without parentheses.

Continuing with undetected errors, the examination revealed that all these errors were semantic in nature. Semantic errors, unlike syntax errors, don't prevent a program from running but lead to unintended results. In the exercise "Exploring Python Data Types", the chatbot failed to identify a semantic error where a wrong data type was assigned to a variable:"

```
integer_var = "5"   #Should be integer, not string
```

In the "Comparing Values with Comparison Operators" exercise, it overlooked errors stemming from the use of incorrect comparison operators:

```
greater_result = num1 >= num2 #Should be num1 > num2
```

and

```
greater_result = num1 < num2   #Should be num1 > num2
```

For both the "Defining and Calling Custom Functions" and "Working with Function Return Values" exercises, the chatbot missed errors related to the omission of function calls:

```
def greet(name):
    print(f"Hello, {name}!")
greet   # Forgot to call the function with parentheses and an argument
```

and

```
def add(num1, num2):
    return num1 + num2
result = add   # Forgot to call the function with parentheses and arguments
```

The observation underscores a significant distinction: syntax errors, which prevent a program from executing, are relatively straightforward to detect and diagnose. In contrast, semantic errors, which do not impede program execution but yield incorrect results, pose a more elusive challenge. Identifying semantic errors necessitates understanding the code's underlying intention, presenting a distinct challenge for chatbots like PythonPal. Nevertheless, the Test Case module addresses this by pinpointing instances where a student's code output diverges from the anticipated outcome. When such discrepancies arise, the module designates the outcome as incorrect, signaling a potential semantic error. Thus, the Test Case module demonstrates its capability to detect semantic errors within the provided test cases.

Furthermore, certain unrecognizable error patterns were noted in exercises like "Practice Basic Arithmetic Operations" and "Comparing Values with Comparison Operators". Although PythonPal detected these syntax errors, they did not correspond to any pre-established patterns in the system. In such scenarios, PythonPal reverts to presenting Python's original error message, which encompasses the error type and the line number of its occurrence.

For instance, in the "Practice Basic Arithmetic Operations" exercise, the system didn't recognize syntax errors caused by a missing assignment operator for variable 'b':

```
b 0
```

For the "Comparing Values with Comparison Operators" exercise, the system overlooked an error due to a missing operator:

```
greater_result = num1 num2
```

The unidentified error patterns underscore areas where PythonPal's error detection capabilities can be enhanced. These oversights are recognized, and there are plans to augment the system to identify such patterns in subsequent versions. This entails refining the pattern-matching capabilities to encompass a wider array of syntax errors, thus elevating PythonPal's comprehensive error detection and explanation performance.

*Conclusion for RQ1*

PythonPal demonstrates proficiency in identifying and elucidating basic syntax errors; however, challenges arise with semantic errors and some atypical syntax issues. The system adeptly flags errors in exercises such as "Variable Creation and Assignment" but overlooks subtle mistakes, particularly those that do not interrupt code execution but yield incorrect outcomes. When faced with unfamiliar error patterns, the system reverts to Python's inherent error messages. In summary, while PythonPal excels in pinpointing and providing feedback on syntax errors, there is room for enhancement in addressing semantic errors and certain unconventional syntax errors. Future development efforts will prioritize refining these areas to achieve a more holistic and precise system for programming education.

With the ability of PythonPal to identify basic syntax errors being acknowledged, and the areas for improvement recognized, attention is now extended to the impact of these technical aspects on the learner's journey. The effectiveness of the system from the perspective of students will be further detailed in the following section.

### D. User Satisfaction Evaluation

This section addresses Research Question 2 (RQ2): How do students perceive PythonPal's ability to comprehend queries, provide accurate feedback, and ensure overall satisfaction in the learning experience? It aims to assess user satisfaction with PythonPal, focusing on its query comprehension and feedback accuracy, to evaluate its effectiveness as an educational tool.

*Participants*

The research population included 21 second-year students from the Data Analytics for Business course. These participants had no previous coding experience, positioning them as suitable subjects for assessing a programming-centric chatbot. Due to various commitments, three students were unable to participate, leading to a final participant count of 18, which is approximately 88.71% of the initial course enrollment. This research utilized a convenient sampling technique, given limited access to other potential participant groups, as the researcher also served as the course instructor. While convenience sampling presents challenges, such as potential bias and limited generalizability, it meets the immediate objectives of this preliminary study, which seeks to understand student perceptions and gather feedback for subsequent enhancements. There are intentions to broaden the research scope in future investigations, aiming to encompass a wider variety of students, potentially from different academic years or institutions, to ensure a more thorough assessment of the chatbot's efficacy.

To enhance the precision and dependability of the feedback, the research project was clearly explained, and the application was demonstrated.

*Data Collection*

The data collection process for this study entailed direct interaction with the PythonPal chatbot. Participants accessed the chatbot and engaged in dialogue, requested tutorials, and completed exercises. Upon exercise completion, the chatbot offered feedback, enabling participants to rectify errors and resubmit their tasks.

Upon completion of this interaction, students were asked to respond to a set of three questions designed to evaluate their experience with the chatbot. The questions were as follows:

1) "How do you rate PythonPal's ability to understand your queries?" (Comprehension)
2) "How do you perceive the accuracy of feedback provided by PythonPal in response to your programming exercises or queries?" (Correctness)
3) "How satisfied are you with your overall experience using PythonPal?" (Satisfaction)

Each question was rated on a Likert scale of 1 to 5, with 1 indicating a negative response (e.g., poor understanding, not at all effective, very dissatisfied) and 5 indicating a positive response (e.g., excellent understanding, extremely effective, very satisfied). In addition to these questions, the survey included an optional comment section where students were encouraged to provide written feedback on their experience. This section aimed to capture more nuanced opinions and suggestions for improvement.

The design of these questions directly targets key aspects of PythonPal's performance, focusing on its capability to understand student queries, deliver accurate feedback, and ensure overall user satisfaction. This approach aims to uncover insights related to Research Question 2 (RQ2), concentrating on user perceptions of PythonPal's performance across these critical dimensions. The correlation among these dimensions stems from their sequential nature in the students 's interaction with PythonPal. The system's ability to understand queries (Comprehension) directly impacts its ability to provide accurate feedback (Correctness), both of which together influence the user's overall satisfaction (Satisfaction).

The survey questions draw from the End-User Computing Satisfaction (EUCS) model by Doll, Xia, and Torkzadeh [44] which has been validated across various contexts, including educational technologies, for assessing user satisfaction with information systems. Specifically, the question on PythonPal's query comprehension aligns with the EUCS's content facet, emphasizing the importance of accurately understanding user input, which results in delivering the correct content. The inquiry into the accuracy of PythonPal's feedback reflects the EUCS's accuracy facet, focusing on the precision of the information provided. Lastly, the overall satisfaction question mirrors the comprehensive assessment approach of the EUCS model, capturing users' cumulative experience with the system. This approach not only acknowledges the foundational framework provided by the EUCS model but also customizes it to explore the distinct interactions users have with PythonPal, focusing on key areas that directly impact educational effectiveness. The adaptation of the EUCS model for our study is informed by previous research [45], [46], [47] that demonstrates its effectiveness within educational settings. These studies validate the EUCS model's utility in exploring user satisfaction dimensions relevant to educational technology, thus substantiating our methodological approach for evaluating PythonPal.

*The evaluation results*

Based on the detailed survey results presented in Table VI, participants provided feedback on the effectiveness of PythonPal across multiple dimensions. In the Comprehension category, Python-Pal received a categorical mean of 3.0556 with a standard deviation (SD) of 1.0556. The verbal interpretation rated this dimension as "Average", suggesting that PythonPal's capability to understand user queries is satisfactory, though there's potential for enhancement. In the Correctness category, the chatbot achieved a mean score of 3.2222 and an SD of 1.0007, also earning an "Average" verbal interpretation. This suggests that while PythonPal tends to provide correct feedback, it could benefit from further refinements. For the Satisfaction category, the platform achieved a categorical mean of 3.2778 with an SD of 0.9583. Again, the verbal interpretation for this category is "Average", indicating that users had a generally satisfactory experience with the platform. It's worth noting that the survey questions had high internal consistency, as evidenced by a Cronbach's α value of 0.9236.

TABLE VI
USER EVALUATION RESULTS

| Category | Categorical Mean | SD | Verbal Interpretation |
|---|---|---|---|
| Comprehension | 3.0556 | 1.0556 | Average |
| Correctness | 3.2222 | 1.0007 | Average |
| Satisfaction | 3.2778 | 0.9583 | Average |

There were 14 comments from students regarding their experience with PythonPal:

1) "Actually, I did not get the answer...The bot responds should be fast... able to respond to relate the question."
2) "I see that PythonPal can give me detailed feedback and enjoy using it."
3) "I accidentally copied pasted the letters...I think the bot's language should be improved."
4) "Very first experience and I enjoyed it."
5) "PythonPal is a little bit slow."
6) "I will be waiting for more coding on PythonPal...I found a little bit issue while using the line bot."
7) "I think the idea of creating Chatbot is great overall... the response was so slow and it didn't give the accurate responses."
8) "It was really satisfying to learn python codes and gets to know how it works."

9) "I think this is a very smart bot...So it can misunderstand easily. That's my overall view."
10) "Respond too late."
11) "Needs to develop the understanding of the commands and the response time. It cannot understand the commands well."
12) "It was quite slow."
13) "It was fun!"
14) "First time learning Python. It was a great experience."

Valuable feedback emerged from student comments, highlighting several key areas of PythonPal:

- Response Time: As noted in comments 1, 5, 7, 10, and 12, students desire faster responses from PythonPal. It's important to note that actual response time includes both PythonPal's processing and additional network latency from the user's device and the LINE app. This can extend the perceived response time, suggesting an area for future optimization to align user experience with system performance.
- Understanding User Inputs: Comments 1, 7, 9, and 11 indicate occasional challenges in PythonPal's understanding of user queries. Improvements in recognizing user intent seem warranted.
- Bot's Language and Tone: The tone used by PythonPal drew attention in comments 3 and 9. Maintaining a friendly, neutral tone can positively impact user experience.
- Positive Feedback: It's encouraging to observe positive remarks in comments 2, 4, 8, 13, and 14. These comments shed light on PythonPal's benefits and the satisfaction it brings to some users.

In summary, the feedback underscores PythonPal's potential and the aspects that resonate with users. At the same time, it outlines areas for refinement. With this feedback in hand, there's a clear roadmap for enhancing both PythonPal's technical capabilities and its user interactions.

*Conclusion for RQ2*

Based on survey data and student comments, PythonPal demonstrates an average capability in understanding user queries, as evidenced by a mean score of 3.0556 in comprehension. Its proficiency in offering accurate feedback is also rated as "Average," with a mean of 3.2222, indicating areas that might benefit from enhancements. The general satisfaction is similarly categorized as "Average" with a score of 3.2778. While the feedback quality is acknowledged positively by students, concerns regarding response speed and the bot's formal tone have been noted. Misunderstandings due to occasional misclassification of user intent have also been observed, confusing some users.

To enhance the user experience, PythonPal will focus on improving intent classification accuracy and response times. In parallel, an investigation will be conducted into how variations in the efficiency of PythonPal's models influence user perceptions and interactions, with the aim of enhancing the chatbot's efficiency and user-friendliness, thereby improving overall satisfaction rates.

## VI. LIMITATIONS

The following discussion addresses the study's limitations and suggests directions for future research. Firstly, the limitation stemming from the lack of a comparative analysis with traditional e-learning methods or a control group curtails the ability to ascertain PythonPal's efficacy. This limitation will be addressed by including a controlled study in future research to measure learning outcomes and user satisfaction, comparing these against conventional e-learning platforms. By conducting this comparative research, the benefits of PythonPal are aimed to be substantiated and its effectiveness as an innovative educational tool validated.

Another limitation pertains to the social dimensions of chatbot-mediated learning. While this study primarily focused on PythonPal's technical strengths—such as its accuracy in error detection and immediacy of feedback—these features indirectly support the overall learning experience by enhancing student engagement and motivation. Several student comments reflect these dynamics. For instance, one student noted, "I see that PythonPal can give me detailed feedback and enjoy using it" (comment 2), highlighting how detailed, timely feedback can foster engagement. Another student remarked, "It was fun!" (comment 13), reflecting the potential for motivation and enjoyment in using the system. Additionally, "First time learning Python. It was a great experience" (comment 14) underscores how positive experiences with PythonPal can contribute to student satisfaction and motivation.

However, despite these positive effects on individual motivation, the role of social interactions was not deeply explored in this study. While personalized and adaptive feedback can significantly boost engagement and motivation, as shown by studies from Marwan et al. [20] and Ayedoun et al [24], this research did not investigate how PythonPal's technical features might further enhance motivation through social learning outcomes. Future research could explore how to build upon these motivational aspects to better support engagement and sustained learning outcomes.

PythonPal's real-time feedback system has already demonstrated several implications for engagement and motivation, aligning with Self-Determination Theory (SDT), which identifies autonomy and competence as key drivers of intrinsic motivation [48]. PythonPal supports these needs by allowing students to control their learning pace (autonomy) and providing immediate feedback on their performance (competence). Several students highlighted this in their feedback, such as, "It was really satisfying to learn Python codes and get to know how it works" (comment 8), illustrating how the system helps build competence through hands-on learning. "Very first experience and I enjoyed it" (comment 4) reflects the positive learning experience fostered by PythonPal. This personalized, immediate feedback fosters engagement and intrinsic motivation. However, while PythonPal effectively supports individual learning, future research could explore how these features influence broader motivational outcomes. Studies, such as those by Yin et al. [22] and by Schouten et al. [49] confirm that chatbot-based formative feedback significantly boosts intrinsic motivation by meeting psychological needs. Personalized feedback also reduces cognitive load [23], enabling students to focus more on applying new concepts, thereby improving engagement and maintaining motivation throughout their learning process.

Future research should explore both the technical and social impacts of PythonPal's features on engagement, motivation, and learning outcomes through targeted surveys and controlled experiments. A pre-test/post-test design could investigate how various aspects of PythonPal, such as feedback timing and specificity, as well as other interactive features, affect student engagement and motivation. Student engagement can be empirically measured through methods such as self-report questionnaires, interaction logs, task completion rates, and time-on-task metrics, while motivation can be assessed using established scales like the Intrinsic Motivation Inventory (IMI) [48] or Self-Regulation Questionnaire (SRQ) [50]. Previous studies have shown the effectiveness of such methods for assessing engagement and motivation [13], [20], [22]. Statistical analyses, such as t-tests and regression models, could help identify which features have the most significant impact on learning outcomes. This comprehensive understanding will refine PythonPal's design, allowing it to better support both technical and social learning outcomes.

In addition to its technical strengths, PythonPal's conversational design as a text-based tutor could also shape how students perceive and engage with social dimensions of learning. Techniques such as personalized scaffolding, which guides students through complex tasks [23], and adaptive feedback systems that adjust based on learner progress, have been shown to significantly enhance motivation and engagement [22]. To foster collaboration, PythonPal could integrate with a multiagent-based system [27], [28] to facilitate peer interaction, creating opportunities for social learning, though this would require significant development as part of a larger-scale project to enhance collaborative learning. Future iterations of PythonPal will continue to evolve based on empirical evidence, supporting both technical mastery and enhanced social learning.

## V. Conclusion

This paper has presented the ongoing development of PythonPal, a Python learning chatbot, designed to address the challenges faced by novice programmers. The system, which is currently a work in progress, aims to provide an interactive and personalized learning experience for students learning programming for the first time. PythonPal incorporates a chatbot interface that interacts with students in English, providing tutorial videos, exercises, and feedback based on their individual learning needs.

This study evaluated PythonPal addressing its performance across multiple dimensions such as intent classification, dialogue quality, and user experience. DistilBERT emerged as the top performer in intent classification with an F1-score of 0.9352, offering high accuracy but at the expense of computational speed, suggesting an avenue for future work to optimize performance without compromising accuracy. In terms of conversation generation, BlenderBot proved to be the optimal choice based on its superior METEOR score of 0.1504, balancing high-quality conversations with acceptable response times. As for the user interface, PythonPal employs Distil-BERT to handle a wide variety of student interactions seamlessly.

When evaluating PythonPal's efficacy in coding education (RQ1), the system showed strength in detecting syntax errors but exhibited weaknesses in identifying semantic and some uncommon syntax errors. User feedback (RQ2) pointed to the system's good capability in understanding queries but highlighted areas for improvement, including the accuracy of feedback and response speed. The findings provide valuable insights for future development to make PythonPal a more comprehensive, efficient, and user-friendly platform for programming education.

The next phase of PythonPal's development will be dedicated to technical enhancements to strengthen its foundational capabilities. Key focuses include improving the accuracy of intent classification and optimizing system response times to ensure faster and more effective feedback delivery. Refinements to the Tutorial module will aim at superior content organization, accessibility, and the introduction of a 'Precision Video Playback' feature, leveraging Natural Language Processing to enhance user learning experiences directly. Strategies like parallel processing and server scaling are also under consideration to achieve these goals, intending to elevate PythonPal into a more efficient, responsive, and user-centered educational tool.

While technical improvements remain the primary focus, future evaluations will examine PythonPal's broader educational impact, including comparisons with traditional e-learning methods and its potential to enhance social interactions. Controlled experiments will assess learning outcomes, user satisfaction, and the effect of its features on engagement and motivation. Future work will focus on integrating personalized scaffolding and adaptive feedback to improve student engagement, providing real-time guidance tailored to individual needs.

In the long term, PythonPal has the potential to contribute to programming education by offering scalable, personalized learning solutions. With its ability to provide real-time, context-specific support, PythonPal could enhance both technical mastery and student engagement. As PythonPal continues to evolve, future iterations—integrating advanced AI, personalized scaffolding, and adaptive feedback—could further improve the accessibility and effectiveness of programming education.

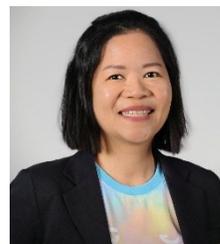

**Sirinda Palahan** received the Ph.D. degree in Computer Science and Engineering from Pennsylvania State University, University Park, PA, USA, in 2014. She is currently an Assistant Professor with the Department of Information and Communication Technology, School of Science and Technology, University of the Thai Chamber of Commerce, Bangkok, Thailand. She serves as both a university educator and corporate trainer in data analytics. Her research interests include business analytics, with focus on predictive modeling and data management for organizations; natural language processing, speech analysis, and sentiment analysis; and educational technology applications, emphasizing learning analytics and intelligent tutoring systems.